\documentclass[prd,twocolumn,preprintnumbers,nofootinbib]{revtex4}
\usepackage[a4paper, hdivide={1.91cm,,1.165cm}, vdivide={1.83cm,,3.1cm}]{geometry}

\usepackage{amsmath,amssymb}
\usepackage{graphicx,multirow}
\usepackage{color}
\usepackage{units}
\usepackage[hyperfootnotes=false]{hyperref}

\def \L {\mathcal{L}} 

\newcommand{\hc}{\ensuremath{\text{h.c.}}}
\newcommand{\matrixx}[1]{\begin{pmatrix} #1 \end{pmatrix}} 
\newcommand{\BR}{\mathrm{BR}}

\def \epsilon {\varepsilon} 

\allowdisplaybreaks

\begin{document}

\preprint{ULB-TH/17-19, PSI-PR-17-15, ZH-TH 27/17}

\title{Large $h\to b s$ in generic two-Higgs-doublet models}

\author{Andreas Crivellin}
\email{andreas.crivellin@cern.ch}
\affiliation{Paul Scherrer Institut, CH--5232 Villigen PSI, Switzerland}

\author{Julian Heeck}
\email{julian.heeck@ulb.ac.be}
\affiliation{Service de Physique Th\'eorique, Universit\'e Libre de Bruxelles, Boulevard du Triomphe, CP225, 1050 Brussels, Belgium}

\author{Dario M\"uller}
\email{dario.mueller@psi.ch}
\affiliation{Paul Scherrer Institut, CH--5232 Villigen PSI, Switzerland}
\affiliation{Physik-Institut, Universit\"at Z\"urich,
	Winterthurerstrasse 190, CH-8057 Z\"urich, Switzerland}

\hypersetup{
pdftitle={Large h to bs in generic two-Higgs-doublet models},
pdfauthor={Andreas Crivellin, Julian Heeck, Dario M\"uller}}


\begin{abstract}
We investigate the possible size of $h\to bs$ in two-Higgs-doublet models with generic Yukawa couplings. Even though the corresponding rates are in general expected to be small due to the indirect constraints from $B_s\to\mu^{+}\mu^{-}$ and $B_s$--$\overline{B}_s$ mixing, we find regions in parameter space where $h\to bs$ can have a sizable branching ratio well above 10\%. This requires a tuning of the neutral scalar masses and their couplings to muons, but then all additional constraints such as $B\to X_s \gamma$, $(g-2)_\mu$, and $h\to\mu^{+}\mu^{-}$ are satisfied. In this case, $h\to bs$ can be a relevant background in $h\to b\bar{b}$ searches and vice versa due to the imperfect $b$-tagging purity. Furthermore, if $h\to bs$ is sizeable, one expects two more scalar resonances in the proximity of $m_h$. We briefly comment on other flavour violating Higgs decays and on the 95 GeV $\gamma\gamma$ resonance within generic two-Higgs-doublet models.
\end{abstract}

\maketitle


\section{Introduction}

The possibility of flavour-changing decays of the Brout--Englert--Higgs boson $h$ (Higgs for short in the following) has been discussed for a long time as a possible signal for physics beyond the Standard Model (SM)~\cite{McWilliams:1980kj,Shanker:1981mj,Han:2000jz,Giudice:2008uua,Goudelis:2011un,Harnik:2012pb,Blankenburg:2012ex}. Indirect constraints on these couplings come from flavour-changing neutral-current observables. In many analyses one follows an effective-field-theory approach in which one assumes that only the couplings of the SM-like Higgs to fermions are modified and derives constraints on these couplings from low-energy processes~\cite{Harnik:2012pb,Blankenburg:2012ex}. This leads one to conclude that no flavour-changing Higgs decays can be observable at the LHC, with the possible exception of $h\to \tau e$ and $h\to\tau\mu$~\cite{Blankenburg:2012ex,Harnik:2012pb}. This is a dangerous conclusion because the very existence of flavour-changing Higgs couplings in a renormalizable SM extension implies additional states which posses flavour-changing couplings as well. The indirect constraints from flavour-changing neutral currents and rare decays are thus inherently model-dependent and can be decoupled from Higgs decays. This generically involves finetuning of the mass spectrum and couplings of the additional states, but opens the way for some new channels to look for physics beyond the SM.

In this article we will study the arguably simplest SM extension that can lead to flavour-changing couplings of the SM-like Higgs: the two-Higgs-Doublet Model (2HDM) with generic Yukawa couplings, i.e.~type III.\footnote{Similar analyses were performed in the MSSM~\cite{Arhrib:2006vy,Barenboim:2015fya,Gomez:2015duj}, also with additional vector-like fermions~\cite{Ibrahim:2017kay} and in 2HDMs with of type I and II~\cite{Arhrib:2004xu}, in aligned 2HDMs~\cite{Gori:2017qwg} as well as in Branco--Grimus--Lavoura~\cite{Branco:1996bq} 2HDMs~\cite{Botella:2015hoa} and Zee models~\cite{Herrero-Garcia:2017xdu}. The correlations between $h\to bs$ and $B_s\to\mu^+\mu^-$ were considered in Ref.~\cite{Chiang:2017etj}.} After computing the effects in $B_s$--$\overline{B}_s$ mixing, $B_s\to\mu^+\mu^-$ and $b\to s\gamma$, we identify regions of parameter space that can lead to sizable decay rates of $h\to bs$ (upwards of 10\%) which are potentially observable at the LHC, hopefully motivating dedicated searches. This is particularly relevant now that the largest Higgs decay mode, $h\to b\bar{b}$, has finally been observed~\cite{ATLAS:2017bic,Sirunyan:2017elk}, rendering it background for $h\to bs$.
While not the focus of our work, we stress that the additional neutral states ($H$ or $A$) can easily have even larger flavour-violating branching ratios, so general resonance searches for $bs$ final states are encouraged as well.

The rest of this article is structured as follows: in Sec.~\ref{sec:typeIII} we set up our 2HDM notation. In Sec.~\ref{sec:observables} we discuss the main observables that could invalidate large $h\to bs$ rates and identify ways to circumvent their constraints. Sec.~\ref{sec:collider} deals with direct searches for the new scalars at colliders, pointing out their main production and decay channels. We comment on different choices of bases for the 2HDM in Sec.~\ref{sec:different_basis}. Finally, we conclude in Sec.~\ref{sec:conclusion} and provide an outlook for other rare Higgs decays.
Appendix~\ref{sec:b_to_s_gamma_formulae} provides one-loop formulae relevant for $b\to s\gamma$.

\section{Type-III 2HDM}
\label{sec:typeIII}

Our starting point is the 2HDM with generic couplings to fermions (type III) and a CP conserving scalar potential~\cite{Branco:2011iw}. In the Higgs basis~\cite{Georgi:1978ri,Lavoura:1994fv,Botella:1994cs} in which only one doublet acquires a vacuum expectation value (using notation close to Ref.~\cite{Davidson:2016utf}) we have                                 
\begin{align}
\Phi_1 =\matrixx{ G^+ \\ \frac{v+H_1^0+i G^0}{\sqrt{2}}}, &&
\Phi_2 = \matrixx{H^+ \\ \frac{H_2^0 + i A}{\sqrt{2}}},
\end{align}
with $v\simeq \unit[246]{GeV}$, the Goldstone bosons $G^{0,+}$, and the physical CP-odd scalar $A$. Assuming that CP is conserved in the scalar potential, the CP-even mass eigenstates are
\begin{align}
h &= H_1^0 \sin (\beta-\alpha) + H_2^0 \cos (\beta-\alpha)\,,\\
H &= H_1^0 \cos (\beta-\alpha) - H_2^0 \sin (\beta-\alpha)\,,
\end{align}
where we defined the mixing angle as $\beta-\alpha$ for easier comparison with the well-known type-I/II/X/Y 2HDM. We will abbreviate $s_{\beta\alpha}\equiv \sin (\beta-\alpha)$, $c_{\beta\alpha}\equiv \cos (\beta-\alpha)$, and $t_{\beta\alpha}\equiv \tan(\beta-\alpha)$ below. 

In the physical basis with diagonal fermion mass matrices the Yukawa couplings are given by
\begin{align}
\begin{split}
\hspace{-1ex}-\L_Y &= \sum_{f=u,d,\ell}\left[\overline{f} \left( y^f s_{\beta\alpha} + (\epsilon^f P_R + {\epsilon^f}^\dagger P_L) c_{\beta\alpha}\right) f \, h\right.\\
&\quad+ \overline{f} \left( y^f c_{\beta\alpha} - (\epsilon^f P_R + {\epsilon^f}^\dagger P_L) s_{\beta\alpha}\right) f \, H\\
&\quad+ \left. i \eta_f\,\overline{f} \left(\epsilon^f P_R - {\epsilon^f}^\dagger P_L\right) f \, A\right]\\
&\quad+ \sqrt{2}\left[ \overline{u}\left( V\epsilon^d P_R - {\epsilon^u}^\dagger V P_L\right) d\, H^+ +\hc\right]\\
&\quad+ \sqrt{2}\left[ \overline{\nu}\left( \epsilon^\ell P_R \right) \ell\, H^+ +\hc\right] ,
\end{split}
\label{eq:yukawas}
\end{align}
where $\eta_{d,\ell} = 1 = -\eta_u$ and $V$ is the Cabibbo--Kobayashi--Maskawa (CKM) matrix. $(y^f)_{ij} = \delta_{ij} m_j^f/v$ are the standard (diagonal) SM Yukawa couplings, while $\epsilon^{u,d,\ell}$ are arbitrary complex $3\times 3$ matrices in flavour space. Off-diagonal elements in $\epsilon^f$ lead to flavour-changing Higgs couplings. For our channel of interest, $h\to \overline{b}s + b\overline{s}$, we have
\begin{align}
\Gamma (h\to bs) \simeq \frac{3 c_{\beta\alpha}^2 m_h}{8\pi}  (|\epsilon_{23}^d|^2+|\epsilon_{32}^d|^2) \left(1-\frac{m_b^2}{m_h^2}\right)^2 .
\end{align}
Note that this expression is valid at tree level; next-to-leading order QCD corrections might increase the decay rate by $10$--$20\%$~\cite{Barenboim:2015fya}. However, since we are interested in an order of magnitude estimate, such corrections are not of particular importance here. The resulting branching ratio is then
\begin{align}
\BR(h\to bs) = \frac{\Gamma(h\to bs)}{\Gamma (h\to bs) + s_{\beta\alpha}^2 \Gamma_\mathrm{SM}}\,,
\end{align}
with $\Gamma_\mathrm{SM}\simeq \unit[4.1]{MeV}$ and assuming all $\epsilon^{u,d,\ell}$ to be zero, except of course those for $h\to bs$. Note that a branching ratio of $h\to b s$ of $1\%$ ($10\%$) requires $\epsilon^d_{23,32}$ couplings of order $0.02$ ($0.06$), assuming $c_{\beta\alpha} = 0.1$.

So far no searches for $h\to bs$ have been performed, making it difficult to assess the sensitivity. The channel $h\to b\bar{b}$, which has a large SM branching ratio of $58\%$, has only recently been observed~\cite{ATLAS:2017bic,Sirunyan:2017elk} despite its better $b$-tagging possibilities compared to $h\to bs$.
Nevertheless, we can obtain a model-independent limit on $\Gamma (h\to bs)$ of $\unit[1.1]{GeV}$~\cite{Sirunyan:2017exp}, corresponding roughly to the CMS energy resolution. This is still almost three orders of magnitude above the SM value $\Gamma_\mathrm{SM}$, and thus still allows for $\BR (h\to bs)\sim 1$. A more intricate upper limit on the Higgs width can be obtained by comparing on- and off-shell cross sections, as proposed in Ref.~\cite{Caola:2013yja}. A recent CMS analysis of run-1 data along these lines obtains $\Gamma_h <\unit[13]{MeV}$~\cite{Khachatryan:2016ctc}. While it cannot be claimed to be a model-independent limit~\cite{Englert:2014aca}, it should hold true in our scenario with $c_{\beta\alpha}\ll 1$, seeing as $h$ becomes arbitrarily SM-like. Naively applying $\Gamma_h <\unit[13]{MeV}$ on our model and using $c_{\beta\alpha}\leq 0.55$ as a very conservative bound (see below), this implies $\BR (h\to bs) \lesssim 78\%$; for $c_{\beta\alpha}\ll 1$ the limit is $\BR (h\to bs) \lesssim 68\%$. This is obviously still very large and can most likely be improved by a direct search for $h\to bs$. We will use this as a conservative limit in the following.

Stronger limits can be obtained from global fits to observed Higgs production and decay channels, seeing as a large $\Gamma (h\to bs)$ would reduce all measured Higgs branching ratios and hence require a larger production cross section to obtain the same rates. An analysis of this type with LHC run-1 data was performed in Ref.~\cite{Khachatryan:2016vau} and lead to the $95\%$~C.L.~limit $\BR (h\to \text{new}) < 34\%$ on any new decay channels, including $bs$. This is a factor of two stronger than the limit from the Higgs width, in part because it is based on a combination of ATLAS and CMS data and makes use of more search channels. We will also show this limit in the following, but stress that it should be taken with a grain of salt; global-fit limits  are very indirect and depend strongly on the assumptions one puts in. With the many parameters available in a type-III 2HDM, it is conceivable that the limit could be weakened by increasing some parameters relevant to Higgs production. A dedicated search for $h\to bs$ will yield far more direct constraints and should always be preferred to global-fit limits.

The goal of our article is to show that a sizable branching ratio for $h\to bs$ is possible, even up to the conservative limit of $68\%$. To simplify the analysis we will set as many entries of $\epsilon^f$ to zero as possible, i.e.~$\epsilon^{u,d,\ell}_{ij} = 0$ is the starting point of our investigation. In this limit, we can obtain bounds on the masses and on the mixing angle $\beta-\alpha$ by comparison with the type-I 2HDM (in the limit $\tan(\beta) \to \infty$, i.e.~$\beta\to \pi/2$, identifying our $c_{\beta\alpha}$ with the type-I $\sin(\alpha) = \cos (\pi/2 -\alpha)$). This gives the rather weak bound $|c_{\beta\alpha}| \lesssim 0.55$ from LHC run-1 Higgs measurements~\cite{Dorsch:2016tab,Marcellini:2017nwk}. 
In the limit $c_{\beta\alpha}\to 0$, the new scalars become completely fermiophobic and the model resembles the Inert Higgs Doublet (IDM), with a $\mathbb{Z}_2$ symmetry that only allows the new scalars to be produced in pairs. This $\mathbb{Z}_2$ is of course broken in the scalar potential and by $c_{\beta\alpha}\neq 0$, but it allows us to use well-known limits on IDM. In particular, LEP constraints on the $Z$ and $W$ widths approximately require
\begin{align}
m_A + m_H\geq m_Z\,, &&
m_{H^+} + m_{A,H} \geq m_W\,,
\end{align}
while LEP-II excludes $m_{H^+}<\unit[70]{GeV}$ and also restricts the $m_A$--$m_H$ parameter space~\cite{Lundstrom:2008ai}. Additional bounds come from LHC searches, which most importantly constrain the masses below $m_h/2$~\cite{Belanger:2015kga,Datta:2016nfz}.
The Peskin--Takeuchi parameters $S$ and $T$ also provide constraints, unless the mass spectrum satisfies $m_A\simeq m_{H^+}$ (for $\Delta T\simeq 0$) and $m_A\simeq m_H\simeq m_{H^+}$ (for $\Delta S\simeq 0$)~\cite{Barbieri:2006dq,Haber:2010bw,Arhrib:2012ia}.
All in all, the fermiophobic limit still allows for new-scalar masses around $\unit[100]{GeV}$, depending on the hierarchy. Turning on the mixing angle $\beta-\alpha$ will significantly affect the limits on $m_H$ as it opens up gluon fusion, diphoton decay, etc., to be discussed below.

\section{Observables}
\label{sec:observables}

Since we are interested in $h\to bs$ we will use the ansatz
\begin{align}
\epsilon^d = \matrixx{0 & 0 & 0\\ 0 & 0 & \epsilon_{23}^d \\ 0 & \epsilon_{32}^d & 0} ,  && \epsilon^\ell = \matrixx{0 & 0 & 0\\ 0 & \epsilon_{\mu\mu}^\ell & 0 \\ 0 & 0 & 0} , &&
\epsilon^u =  0\,, 
\label{eq:epsilon_ansatz}
\end{align}
where in addition to $\epsilon_{23,32}^d$ we also allow for non-zero values of $\epsilon_{\mu\mu}^\ell$ because this entry is important for $B_s\to\mu^+\mu^-$. In addition to $B_s\to\mu^+\mu^-$, the most relevant constraints originate from $B_s$--$\overline{B}_s$ mixing and $B \to X_s \gamma$. 

These channels were also discussed in the MSSM (i.e.~type-II 2HDM), where the $h\to bs$ branching ratio was found to be tiny~\cite{Barenboim:2015fya,Gomez:2015duj}. Here it is important to discuss the difference of our analysis to the MSSM. Even though at the loop-level non-decoupling effects in the MSSM induce non-holomorphic Higgs couplings~\cite{Hempfling:1993kv,Hall:1993gn,Carena:1994bv,Noth:2008tw,Hofer:2009xb,Crivellin:2010er,Crivellin:2011jt,Crivellin:2012zz} (making it a type-III 2HDM), these effects are only corrections to the type-II structure. Therefore, the strong bounds from direct LHC searches for additional Higgs bosons as well as the stringent bounds from $b\to s\gamma$ on the charged Higgs mass of around \unit[570]{GeV} apply~\cite{Misiak:2017bgg}. Furthermore, in the MSSM the angle $\alpha$ is directly related to $m_A\simeq m_{H^+}$, rendering it small and further suppressing $h\to bs$.

\subsection{\texorpdfstring{$B_s$--$\overline{B}_s$}{Bs-Bsbar} mixing}

\begin{figure}[t]
	\includegraphics[width=0.47\textwidth]{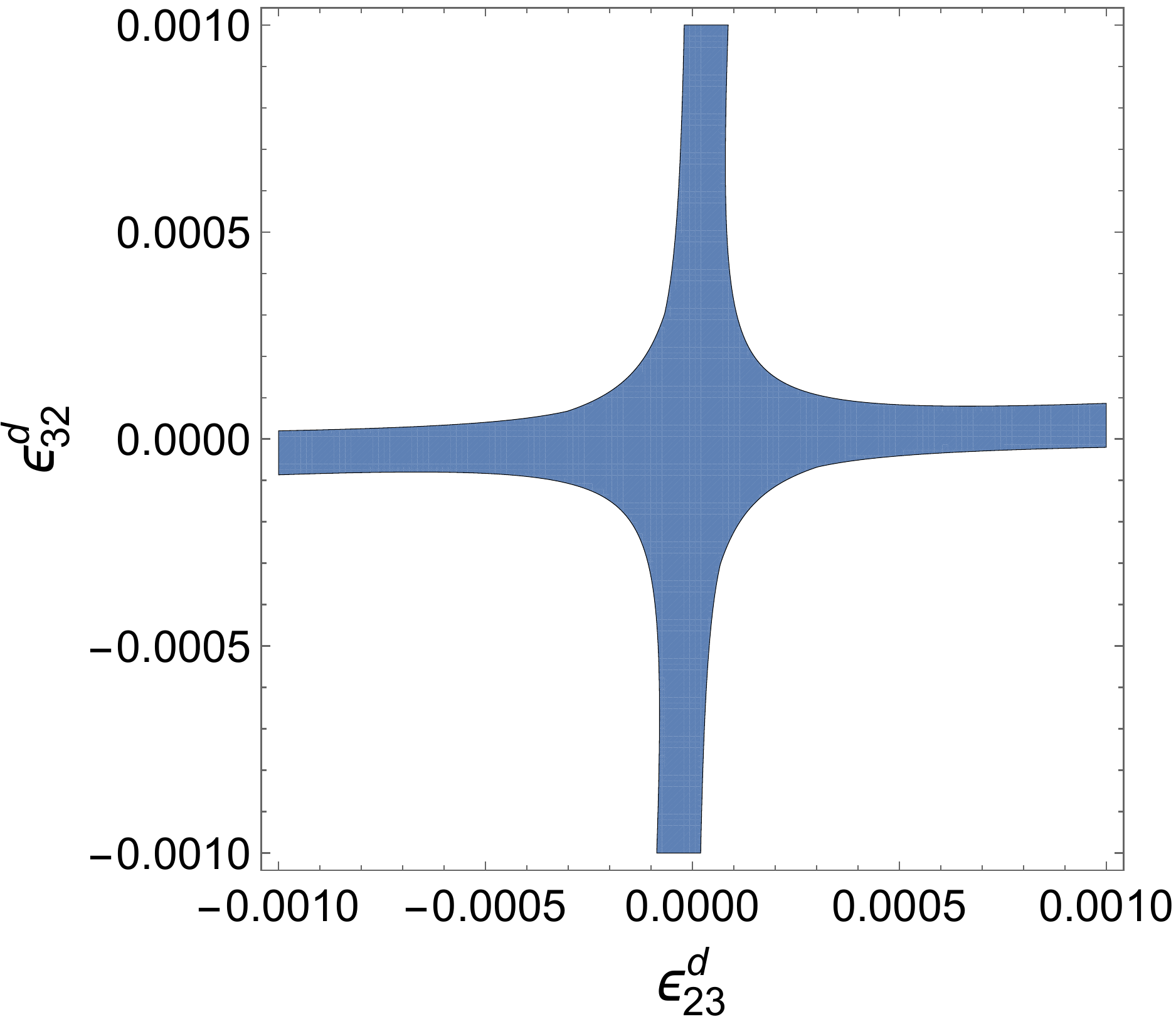}
	\caption{Allowed regions in the $\epsilon^d_{23}$--$\epsilon^d_{32}$ plane for $m_H=\unit[150]{GeV}$ and $c_{\beta\alpha}=0.1$, requiring that the 2HDM contribution to $B_s$--$\overline{B}_s$ mixing should not exceed 10\% compared to the SM which is of the order of the uncertainty in the lattice calculation of the matrix elements. Here we scanned over $m_A$ from 100 to \unit[200]{GeV}. Note that the dependence on $c_{\beta\alpha}$ is very weak. As one can see, in order to get potentially large effects in $h\to b s$, either $\epsilon^d_{23}$ or $\epsilon^d_{32}$ must be very small.	}
	\label{fig:bsmixing}
\end{figure}

The $\Delta F = 2$ process $B_s$--$\overline{B}_s$ mixing is unavoidably modified already at tree-level if $h\to bs$ has a non-vanishing rate. To describe this process we use the effective Hamiltonian (see for example~\cite{Becirevic:2001jj})
\begin{align}
\mathcal{H}_\mathrm{eff} = \sum_{j=1}^5 C_j O_j + \sum_{j=1}^3 C_j' O_j' +\hc\,,
\end{align}
where non-vanishing Wilson coefficients are generated for the three operators
\begin{align}
O_2^{(\prime)} &\equiv (\overline{s}_A  P_{L,(R)} b_A)(\overline{s}_B P_{L,(R)} b_B)\,,\\
O_4 &\equiv (\overline{s}_A  P_L b_A)(\overline{s}_B P_R b_B)\,,
\end{align}
with $A$ and $B$ being colour indices. At tree level, we obtain the Wilson coefficients~\cite{Crivellin:2013wna}
\begin{align}
C_2 &= -\frac{(\epsilon^{d\,\star}_{32})^2}{2} \left[ \frac{c_{\beta\alpha}^2}{m_h^2}+\frac{s_{\beta\alpha}^2}{m_H^2}-\frac{1}{m_A^2}\right],\\
C_2' &= -\frac{(\epsilon^{d}_{23})^2}{2} \left[ \frac{c_{\beta\alpha}^2}{m_h^2}+\frac{s_{\beta\alpha}^2}{m_H^2}-\frac{1}{m_A^2}\right],\\
C_4 &= -(\epsilon^{d\,\star}_{32}\epsilon^{d}_{23}) \left[ \frac{c_{\beta\alpha}^2}{m_h^2}+\frac{s_{\beta\alpha}^2}{m_H^2}+\frac{1}{m_A^2}\right].
\end{align}
Computing the $B_s$--$\overline{B}_s$ mass difference by inserting the matrix elements together with the corresponding bag factor and taking into account the renormalization group evolution~\cite{Becirevic:2001jj}, we show the result in Fig.~\ref{fig:bsmixing}. Here, one can see that $\epsilon^{d}_{23}$ ($\epsilon^{d}_{32}$) can only be sizable if $\epsilon^{d}_{32}$ ($\epsilon^{d}_{23}$) is close to zero. In fact, one can avoid \emph{any} effect in $B_s$--$\overline{B}_s$ mixing by setting 
\begin{align}
m_A = \frac{m_h m_H}{\sqrt{m_h^2 s_{\beta\alpha}^2+m_H^2 c_{\beta\alpha}^2}}\,, && \epsilon^{d\,\star}_{32}\epsilon^{d}_{23} = 0\,.
\label{eq:mA_kills_Bmix}
\end{align}
This in particular implies that $m_A$ is between $m_h$ and $m_H$, so neither the heaviest nor the lightest neutral scalar.
Even with all new-physics Wilson coefficients vanishing at tree level, loop contributions, including those with $H^+$, will generate additional contributions. However, since all contributions interfere, this effect is significantly suppressed compared to the tree-level exchange and can always be cancelled by a small modification of Eq.~\eqref{eq:mA_kills_Bmix}.
We can hence eliminate any new-physics effect in $B_s$--$\overline{B}_s$ mixing using Eq.~\eqref{eq:mA_kills_Bmix} while keeping either $\epsilon^{d}_{23}$ or $\epsilon^{d}_{32}$ large.

\begin{figure*}[t]
	\includegraphics[width=0.42\textwidth]{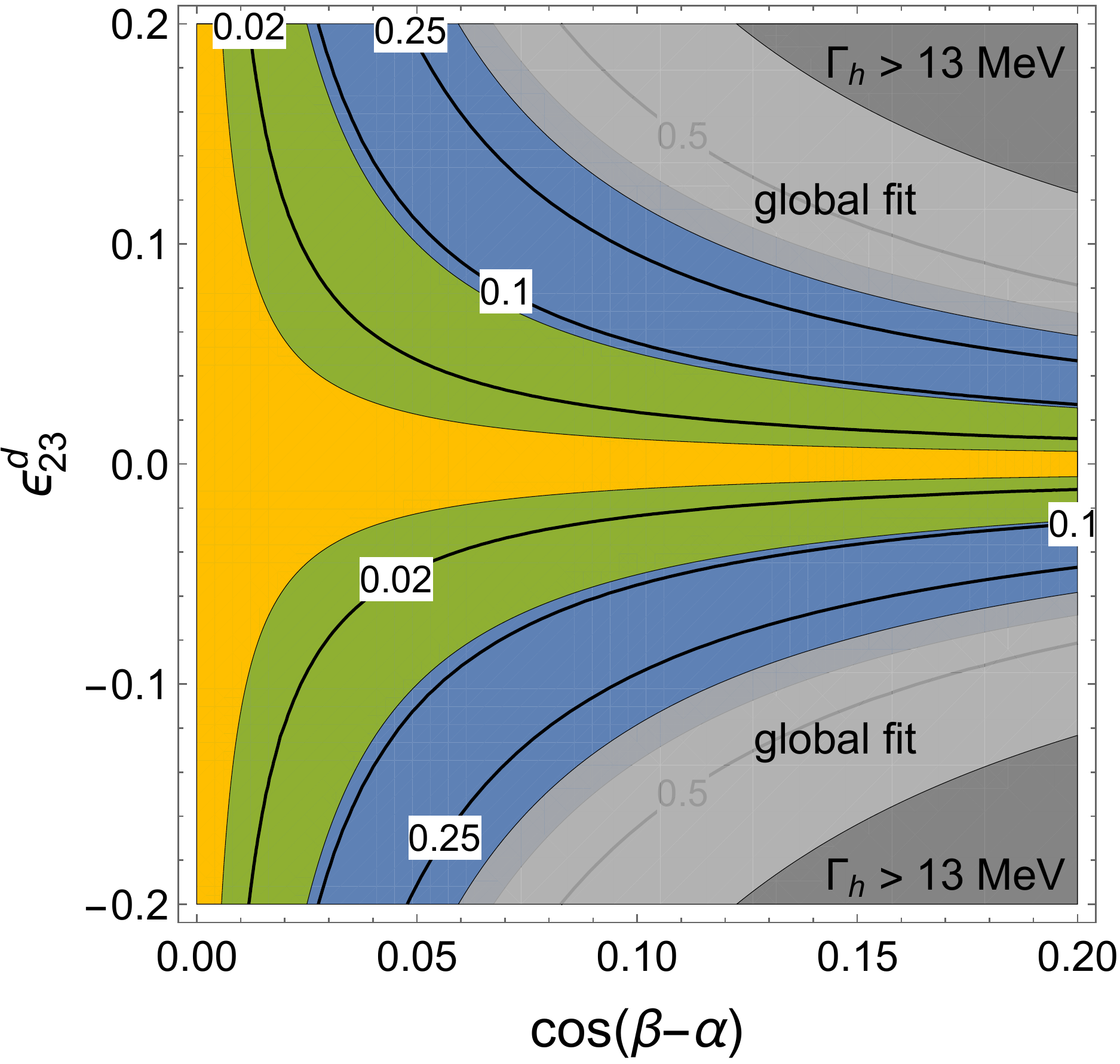}
	\includegraphics[width=0.55\textwidth]{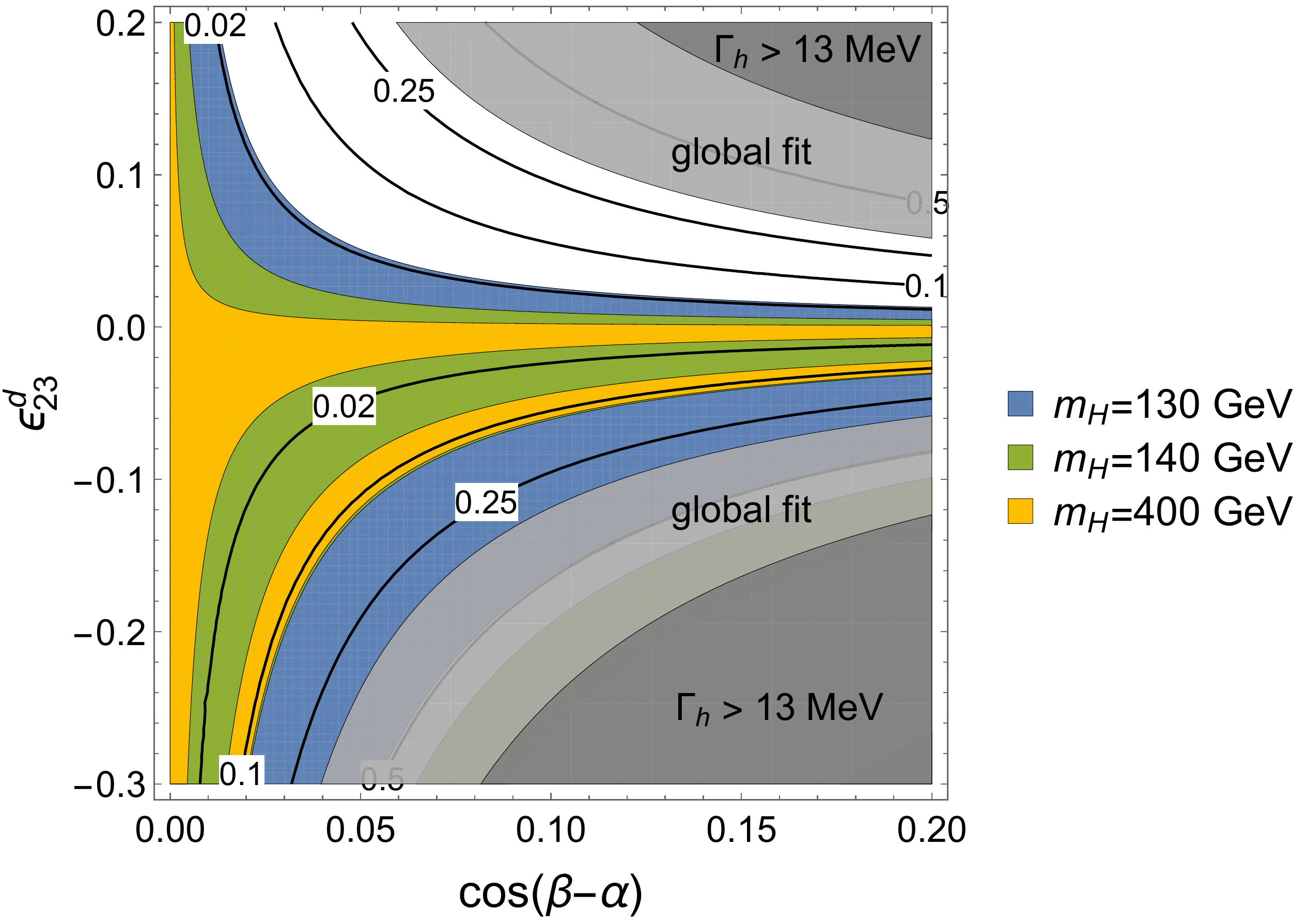}
	\caption{
		$\BR(h\to bs)$ contours and the allowed $2\sigma$ regions from $B_s\to \mu^+\mu^-$. Left: setting all $\epsilon^{f}=0$ except for $\epsilon^d_{23}$, with $m_A$ given by Eq.~\eqref{eq:mA_kills_Bmix}. Right: Same as left plot but with non-zero $\epsilon_{\mu\mu}^\ell$ given by Eq.~\eqref{eq:epsilon_mumu_kills_CS} instead. The darker gray region is excluded by the upper limit on the total decay width of the Higgs of \unit[13]{MeV}~\cite{Khachatryan:2016ctc} and the lighter gray region is excluded by the global-fit constraint $\BR (h\to \rm{anything}) < 34\%$~\cite{Khachatryan:2016vau}.
		}
	\label{fig:bsmumu}
\end{figure*}

\subsection{\texorpdfstring{$B_s \to \mu^+\mu^-$}{Bs to mu mu}}

The $\epsilon^d_{23,32}$ couplings necessary for $h\to bs$ also induce a modification of $B_s \to \mu^+\mu^-$ at tree level, because by construction all three neutral scalars couple to $bs$, and at least two scalars also couple to $\mu^{+}\mu^{-}$.
The effective Hamiltonian takes the form~\cite{Crivellin:2013wna}
\begin{align}
\begin{split}
\mathcal{H}_\mathrm{eff}^{B_s\to\mu\mu} &= -\frac{G_F^2 m_W^2}{\pi^2} \left[C_A O_A + C_S O_S + C_P O_P \right.\\
&\quad\left.+ C_A' O_A' + C_S' O_S' + C_P' O_P'\right]+\hc,
\end{split}
\end{align}
with
\begin{align}
O_A &\equiv (\overline{b} \gamma_\alpha P_L s)(\overline{\mu} \gamma^\alpha \gamma_5 \mu)\,,\\
Q_S &\equiv (\overline{b} P_L s)(\overline{\mu} \mu)\,,\\
Q_P &\equiv (\overline{b} P_L s)(\overline{\mu}\gamma_5 \mu)\,.
\end{align}
$O_X'$ are obtained from $O_X$ by replacing $P_L $ with $P_R$. The branching ratio then reads~\cite{Crivellin:2013wna}
\begin{align}
&\BR (B_s\to\mu^+\mu^-) = \frac{G_F^4 m_W^4}{8\pi^5} \sqrt{1- \frac{4m_\mu^2}{M_{B_s}^2}} M_{B_s} f_{B_s}^2 m_\mu^2 \tau_{B_s} \nonumber\\
&\qquad\times \left[\left|\frac{M_{B_s}^2 (C_P-C_P')}{2 m_\mu (m_b+m_s)} - (C_A-C_A')\right|^2\right. \nonumber\\
&\qquad\qquad+\left.\left|\frac{M_{B_s}^2 (C_S-C_S')}{2 m_\mu (m_b+m_s)}\right|^2 \left(1-\frac{4m_\mu^2}{M_{B_s}^2}\right)\right] \,,
\end{align}
experimentally determined to be $\left(2.8^{+0.7}_{-0.6}\right)\times 10^{-9}$~\cite{CMS:2014xfa}.
The SM yields only one non-zero Wilson coefficient, $C_A^\text{SM} \sim -V_{ts} V_{tb}^*$, while our neutral scalars induce
\begin{align}
\begin{split}
C_S-&C_S' = \frac{\pi^2}{G_F^2 m_W^2}\left[ \frac{i}{m_A^2} \Im (\epsilon_{\mu\mu}^\ell)\left(\epsilon_{23}^{d\,\star}+\epsilon_{32}^d\right)\right.\\
&+\frac{c_{\beta\alpha}}{m_h^2}\left( y_{\mu\mu}^\ell s_{\beta\alpha} + \Re (\epsilon_{\mu\mu}^\ell) c_{\beta\alpha}\right)\left(\epsilon_{23}^{d\,\star}-\epsilon_{32}^d\right)\\
&\left. -\frac{s_{\beta\alpha}}{m_H^2}\left( y_{\mu\mu}^\ell c_{\beta\alpha} - \Re (\epsilon_{\mu\mu}^\ell) s_{\beta\alpha}\right)\left(\epsilon_{23}^{d\,\star}-\epsilon_{32}^d\right)\right],
\end{split}\\
\begin{split}
C_P-&C_P' = \frac{\pi^2}{G_F^2 m_W^2}\left[ \frac{1}{m_A^2} \Re (\epsilon_{\mu\mu}^\ell) \left(\epsilon_{23}^{d\,\star}+\epsilon_{32}^d\right)\right.\\
&\left. + i\left(\frac{c_{\beta\alpha}^2}{m_h^2} +\frac{s_{\beta\alpha}^2}{m_H^2} \right)\Im (\epsilon_{\mu\mu}^\ell)\left(\epsilon_{23}^{d\,\star}-\epsilon_{32}^d\right)\right].
\end{split}
\end{align}
First of all note that one cannot avoid effects here by setting $\epsilon_{32}^d=\pm \epsilon_{23}^{d\,\star}$ due to the constraints from $B_s$--$\overline{B}_s$ mixing. Adjusting $\epsilon^\ell_{\mu\mu}$ allows one to eliminate the muon coupling of at most one of the neutral scalars, leaving the other two contributing to $B_s \to \mu^+\mu^-$ at tree level. Setting for example $\epsilon_{\mu\mu}^\ell = 0$ gives $C_P-C_P'=0$ and $C_S-C_S'\propto (1/m_h^2 - 1/m_H^2)\left(\epsilon_{23}^{d\,\star}-\epsilon_{32}^d\right)$, which can only be made small for $m_H\sim m_h$ with our ansatz from Eq.~\eqref{eq:mA_kills_Bmix}. As can be seen in Fig.~\ref{fig:bsmumu} (left), this is already sufficient to obtain $\BR (h\to bs) = \mathcal{O}(10\%)$ while satisfying the experimental $B_s\to\mu^{+}\mu^{-}$ result within $2\sigma$.

An even better ansatz is to choose a (real) $\epsilon_{\mu\mu}^\ell$ such that $C_S-C_S^\prime = 0$, as this allows for new-physics contributions interfering with the SM Wilson coefficient $C_A$. The required coupling for $C_S-C_S^\prime = 0$ is\footnote{This coupling results in $\BR(h\to\mu^{+}\mu^{-}) \leq \BR(h\to\mu^{+}\mu^{-})_\text{SM}$ for $\unit[90]{GeV}< m_H$, so we automatically evade current LHC limits on this so far unobserved decay mode~\cite{Aaboud:2017ojs}.}
\begin{align}
\epsilon_{\mu\mu}^\ell = \frac{c_{\beta\alpha}s_{\beta\alpha}\left(m_h^2-m_H^2\right) }{c_{\beta\alpha}^2 m_H^2+s_{\beta\alpha}^2 m_h^2 }  y_{\mu\mu}^\ell \,,
\label{eq:epsilon_mumu_kills_CS}
\end{align}
which gives, using also Eq.~\eqref{eq:mA_kills_Bmix},
\begin{align}
C_P-C_P' =\frac{\pi^2 y_{\mu\mu}^\ell c_{\beta\alpha}s_{\beta\alpha}}{G_F^2 m_W^2}\left(\frac{1}{m_H^2}-\frac{1}{m_h^2}\right)  \left(\epsilon_{23}^{d\,\star}+\epsilon_{32}^d\right) .
\label{eq:CP_with_muon_coupling}
\end{align}
The most obvious way to eliminate the new-physics effect here is to choose $m_H=m_h$, which also implies $m_A = m_h$ with Eq.~\eqref{eq:mA_kills_Bmix}. Another possibility is to pick the phase of $\epsilon_{23,32}^{d}$ in such a way that it induces destructive interference with the SM contribution $C_A$, which will soften the limits and allow for larger $h\to bs$, see Fig.~\ref{fig:bsmumu} (right). The largest possible $h\to bs$ values arise when $C_P-C_P'$ destructively interferes with $C_A^\text{SM}$, while keeping $\BR (B_s\to\mu^+\mu^-)$ close to its SM value. Indeed, if we impose the condition
\begin{align}
\frac{M_{B_s}^2 (C_P-C_P')}{2 m_\mu (m_b+m_s)} - C_A^\text{SM} \stackrel{!}{=} + C_A^\text{SM} \,,
\end{align}
then all observables in $B_s\to\mu^{+}\mu^{-}$ will remain exactly at their SM values, as we are effectively just flipping the sign of the SM contribution, which is unphysical (see, for example, Refs.~\cite{Blake:2016olu,Fleischer:2017yox}).
The above relation can be immediately solved for $\epsilon_{23,32}^{d}$
\begin{align}
\epsilon_{23}^{d\,\star}+\epsilon_{32}^d = \frac{4 G_F^2 m_W^2 m_b m_\mu C_A^\text{SM} m_h^2 m_H^2}{\pi^2 c_{\beta\alpha}s_{\beta\alpha} y_{\mu\mu}^\ell M_{B_s}^2 (m_h^2-m_H^2)} \,,
\label{eq:ep23_kills_BsMuMu}
\end{align}
where $m_b$ should now be evaluated at the scale $m_H$ to take the running of $C_P-C_P'$ into account~\cite{Crivellin:2013wna}.

\begin{figure*}[t]
	\includegraphics[width=0.7\textwidth]{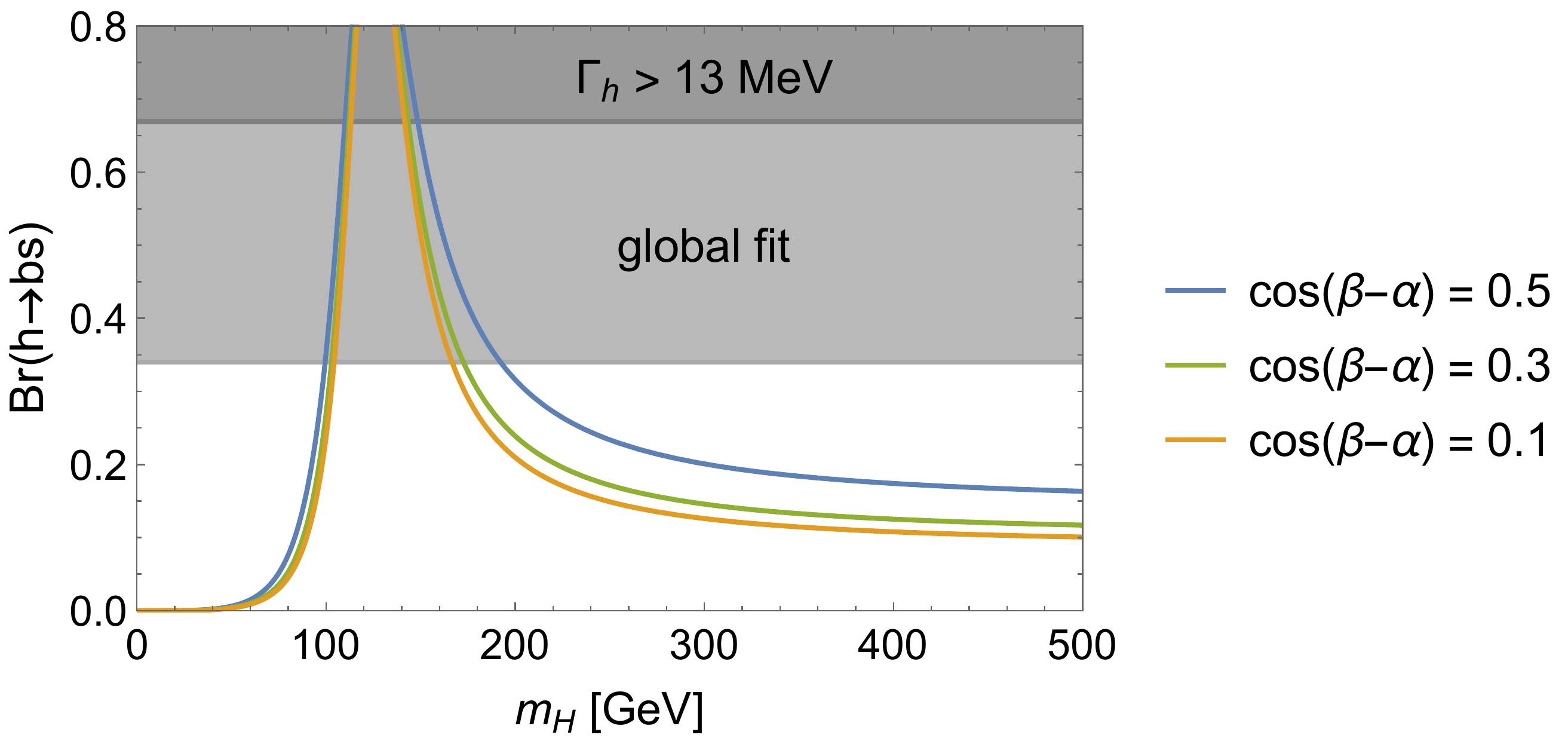}
	\caption{$\BR(h\to bs)$ vs.~$m_H$ for our ansatz from Eqs.~\eqref{eq:mA_kills_Bmix},~\eqref{eq:epsilon_mumu_kills_CS},~\eqref{eq:ep23_kills_BsMuMu} for different values of $c_{\beta\alpha}$. Note that for degenerate masses of the neutral scalars $h\to bs$ is in principle unbounded.}
	\label{fig:mHvsBRbs}
\end{figure*}

To reiterate, choosing masses and couplings according to Eqs.~\eqref{eq:mA_kills_Bmix},~\eqref{eq:epsilon_mumu_kills_CS},~\eqref{eq:ep23_kills_BsMuMu} allows us to keep all $B_s$--$\overline{B}_s$ and $B_s\to\mu^+\mu^-$ observables \emph{at their SM values}, even though $h\to bs$ can be large. The only free relevant parameters left are $c_{\beta\alpha}$ and $m_H$, so we can show $h\to bs$ as a function of $m_H$, see Fig.~\ref{fig:mHvsBRbs}.
As expected, the region $m_H\sim m_h$ allows for the largest $h\to bs$ rates due to the cancellation in $C_P-C_P'$ in Eq.~\eqref{eq:CP_with_muon_coupling}. However, even for $m_h \ll m_H$ and $c_{\beta\alpha} \ll 1$ one can obtain $\BR (h\to bs) \simeq 10\%$.

As an aside, Eq.~\eqref{eq:ep23_kills_BsMuMu} is the only expression so far that depends on the quark flavour, via $C_A^\text{SM} \sim V_{ts} V_{tb}^*$. All our results can thus be easily translated to the case $h\to bd$, with $\Gamma (h\to bd)/\Gamma (h\to bs) \simeq |V_{td}/V_{ts}|^2\simeq 0.05$ in the maximum-cancellation region. A large $h\to bd$ rate above the percent level thus requires $m_H\sim m_A\sim m_h$ if $B_d$--$\bar{B}_d$ mixing and $B\to \mu^{+}\mu^{-}$ are to be kept around their SM values. Hence, larger fine-tuning is needed.

\subsection{\texorpdfstring{$B \to X_s \gamma$}{B to Xs gamma}}

At loop level our new scalars unavoidably modify $B \to X_s \gamma$~\cite{Blake:2016olu}, the relevant one-loop formulae can be found in App.~\ref{sec:b_to_s_gamma_formulae}.\footnote{At two-loop level there can be enhanced Barr--Zee-type contributions~\cite{Barr:1990vd}. However, the maximal enhancement factor is only $m_t/m_b$ (compared to $m_t/m_\mu$ in $(g-2)_\mu$) and including them would not affect our conclusion.} 
Only the Wilson coefficients $C_7$ and $C_8$ are induced in our model.
With our Eqs.~\eqref{eq:mA_kills_Bmix},~\eqref{eq:epsilon_mumu_kills_CS},~\eqref{eq:ep23_kills_BsMuMu}, for the neutral scalars we find that $C_8=- 3 C_7$ depends only on $m_H$ but not on the mixing angle. We can thus predict the size of $C_7$ as a function of $m_H$ (or $\BR(h\to bs)$ with the help of Fig.~\ref{fig:mHvsBRbs}). For $m_H$ in the region of interest for a large $h\to bs$, we find a tiny $|C_7|\simeq 2\times 10^{-3}$ ($2\times 10^{-6}$) for $\epsilon_{32}^d = 0$ ($\epsilon_{23}^d=0$), far below the current limit~\cite{Paul:2016urs}.
$B \to X_s \gamma$ is hence trivially compatible with a large $h\to bs$ in the region of parameter space under study here.

\subsection{\texorpdfstring{$B \to K\ell^{+}\ell^{-}$ and $B_s \to \tau^+\tau^-$}{B to K l l and B to tau tau}}

The decay $B_s\to\mu^{+}\mu^{-}$ is sensitive to the \emph{difference} of the Wilson coefficients $C_{P,S}- C_{P,S}'$, whereas $B\to K\mu^{+}\mu^{-}$ depends on their \emph{sum} $C_{P,S}+ C_{P,S}'$~\cite{Bobeth:2007dw,Becirevic:2012fy,Hiller:2014yaa}. With our ansatz from Eqs.~\eqref{eq:mA_kills_Bmix},~\eqref{eq:epsilon_mumu_kills_CS},~\eqref{eq:ep23_kills_BsMuMu}, we have $C_S = C_S' = 0$ and either $C_P = 0$ or $C_P' = 0$, depending on which $\epsilon^{d}_{23,32}$ we set to zero. We thus unavoidably modify $B\to K\mu^{+}\mu^{-}$ at tree level. Using the results of Ref.~\cite{Becirevic:2012fy}, we checked that this effect is very small, keeping $B\to K\mu^{+}\mu^{-}$ close to the SM value. This means that our model cannot address the observed deviations from the SM prediction in current global fits to $b\to s\mu^+\mu^-$ observables~\cite{Capdevila:2017bsm,Altmannshofer:2017yso,Hurth:2017hxg}.

Similarly, one can expect an effect in $B \to K \tau^+\tau^-$ or $B_s \to \tau^+\tau^-$. Even though the effect is enhanced by $m_\tau/m_\mu$, the very weak experimental bounds on the branching ratio~\cite{TheBaBar:2016xwe,Aaij:2017xqt} (several times $10^{-3}$) do not pose relevant constraints on our parameter space.

\subsection{Anomalous Magnetic Moment of the Muon}

The choice of $\epsilon_{\mu\mu}^\ell$ in Eq.~\eqref{eq:epsilon_mumu_kills_CS} reduces the coupling of $h$ to $\mu\mu$, but enhances the one of $H$ by a factor of few, and also couples $A$ and $H^+$ to muons. As a result, one could expect a modification of $(g-2)_\mu$, an observable that famously deviates from the SM value by around $3\sigma$ and can be explained in 2HDMs~\cite{Iltan:2001nk,Broggio:2014mna,Crivellin:2015hha}. However, the one-loop effect is still suppressed by the small muon mass. In addition, the usually dominant Barr--Zee contributions~\cite{Barr:1990vd} are also not important in our Higgs basis (with a minimal number of free parameters $\epsilon^f$) since the couplings to heavy fermions (top, bottom or tau) are not enhanced for the heavy scalars. Furthermore, $B_s\to\mu^+\mu^-$ prefers nearly degenerate masses for $A$ and $H$, leading to a cancellation in the anomalous magnetic moment of the muon. 

\subsection{ \texorpdfstring{$B^-\to\mu \overline{\nu}$, $D_s^- \to \mu\overline{\nu}$, and $K^- \to \mu\overline{\nu}$}{B- to mu nu, Ds- to mu nu, and K- to mu nu}}

\begin{figure}[t]
	\includegraphics[width=0.48\textwidth]{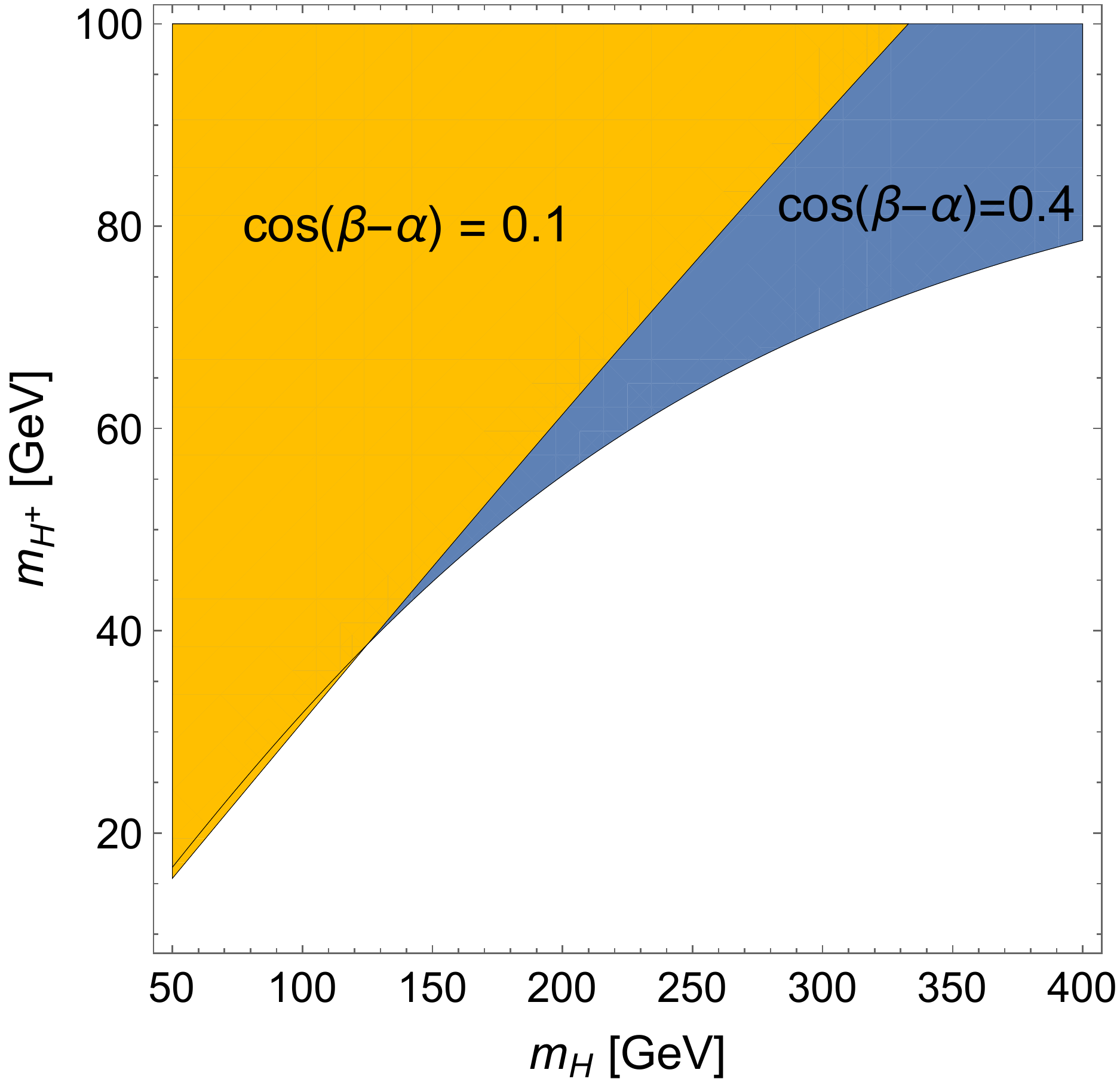}
	\caption{Allowed region from $B\to\mu\nu$ for our ansatz from Eqs.~\eqref{eq:mA_kills_Bmix},~\eqref{eq:epsilon_mumu_kills_CS},~\eqref{eq:ep23_kills_BsMuMu} with  $\epsilon^d_{32} = 0$.	}
	\label{fig:B_to_mu_nu}
\end{figure}

Concerning $H^+$ effects, the best channel is $B^-\to\mu \overline{\nu}$ (assuming $\epsilon^\ell_{\tau\tau}=0$), with the rate
\begin{align}
\frac{\BR (B^-\to \mu\overline{\nu})}{\BR (B^-\to \mu\overline{\nu})_\text{SM}}\simeq \left| 1-\frac{m_{B^-}^2}{m_\mu m_b} \frac{V_{us} \epsilon^d_{23}\epsilon_{\mu\mu}^\ell}{\sqrt{2} G_F V_{ub} m_{H^+}^2}\right|^2 ,
\end{align}
where $m_b$ is again to be evaluated at the scale $m_{H^+}$ to take the running of the Wilson coefficients into account. The predicted SM branching ratio $\BR (B^-\to \mu\overline{\nu})_\text{SM} \simeq 6\times 10^{-7}$ is small and not observed yet, but our new contribution could reach the current upper limit $\BR (B^-\to \mu\overline{\nu})< 10^{-6}$~\cite{Aubert:2009ar}. From Fig.~\ref{fig:B_to_mu_nu} we see that the limits are rather weak and automatically satisfied in the region $m_{H^+}\sim m_H$.

Two other indirect channels are of potential interest: $D_s^- \to \mu\overline{\nu}$ and $K^- \to \mu\overline{\nu}$, the latter of which is suppressed but measured with more accuracy. We have
\begin{align}
\frac{\BR (K^-\to \mu\overline{\nu})}{\BR (K^-\to \mu\overline{\nu})_\text{SM}}\simeq \left| 1-\frac{m_{K^-}^2}{m_\mu m_s} \frac{V_{ub} \epsilon^d_{32}\epsilon_{\mu\mu}^\ell}{\sqrt{2} G_F V_{us} m_{H^+}^2}\right|^2 ,
\end{align}
which gives much weaker bounds than $B^{-}\to \mu\nu$ before.

\section{Collider Constraints}
\label{sec:collider}

Having explored the indirect constraints that come with a large $h\to bs$ decay, let us briefly comment on possible collider searches.

\subsection{Charged Scalar}

The charged scalar has barely played a role in any of the processes discussed so far, thanks to our ansatz for the $\epsilon$ couplings in Eq.~\eqref{eq:epsilon_ansatz} together with Eq.~\eqref{eq:mA_kills_Bmix}. Its mass is hence a more-or-less free parameter, as long as we keep it close enough to $m_{A,H}$ to not induce too large $S$ and $T$ parameters. Let us briefly comment on the $H^+$ phenomenology beyond electroweak precision observables (see also Ref.~\cite{Akeroyd:2016ymd} for a recent review). Aside from gauge couplings, $H^+$ only couples according to Eq.~\eqref{eq:yukawas}:
\begin{align}
-\L_Y &=  \sqrt{2}\left[ \overline{u}\left( V\epsilon^d P_R\right) d + \overline{\nu}\left( \epsilon^\ell P_R \right) \ell \right] H^+ +\hc,
\end{align}
where $\epsilon^\ell$ contains only the non-zero entry $\epsilon_{\mu\mu}^\ell$ and the quark couplings are determined by the matrix
\begin{align}
V\epsilon^d = \matrixx{0 & V_{ub} \epsilon_{32}^d & V_{us} \epsilon_{23}^d \\0 & V_{cb} \epsilon_{32}^d & V_{cs} \epsilon_{23}^d \\0 & V_{tb} \epsilon_{32}^d & V_{ts} \epsilon_{23}^d }.
\end{align}
Since we impose $\epsilon^{d\,\star}_{32}\epsilon^{d}_{23} = 0$ in order to satisfy limits from $B_s$--$\overline{B}_s$ mixing, the $H^+$ couples only either to $b$ or $s$ quarks. In particular, it does not contribute to $b\to s\gamma$.
$\epsilon^d_{23,32}$ is much bigger than the $\epsilon_{\mu\mu}^\ell$ given by Eq.~\eqref{eq:epsilon_mumu_kills_CS} for a sizeable $h\to bs$ rate, so the dominant coupling of $H^+$ is to quarks. If $H^+$ is lighter than the top quark, it can be produced in its decays:

If $\epsilon^d_{32} = 0$, the production channel is $t\to b H^+$, suppressed by $V_{ts}$, followed by $H^+\to \bar{b} c$ with branching ratio $\simeq 1$. This channel has been looked for~\cite{Laurila:2017phk}, with constraints around $|\epsilon^d_{23}|\lesssim 2$ for $m_{H^+}$ between $90$ and $\unit[150]{GeV}$. This is still compatible with a large $h\to bs$ rate as long as $c_{\beta\alpha}$ is not too small ($c_{\beta\alpha}>0.1$). For completeness, we can replace $\epsilon^d_{23}$ directly with the $h\to bs$ branching ratio $\BR_h^{bs}$ to predict
\begin{align}
\frac{\Gamma (t\to b H^+)}{ m_t} \simeq \frac{|V_{ts}|^2 s_{\beta\alpha}^2}{6 c_{\beta\alpha}^2}  \frac{\BR_h^{bs}}{1-\BR_h^{bs}} \frac{\Gamma_h^\text{SM}}{m_h} \left(1-\frac{m_{H^+}^2}{m_t^2}\right)^2 .
\label{eq:top_decay}
\end{align}

If instead  $\epsilon^d_{23} = 0$, the production channel will be $t\to s H^+$ with $\BR(H^+\to \bar{s} c) \simeq 1$. The rate can be obtained from Eq.~\eqref{eq:top_decay} via $V_{ts}\to V_{tb}$, so this channel is enhanced by $|V_{tb}/V_{ts}|^2\simeq 580$ compared to the previous one. Since this final state has only been considered with the production channel $t\to b H^+$~\cite{Aad:2013hla,Khachatryan:2015uua}, we cannot obtain useful limits.

For $H^+$ masses above the top mass the typical search channel is $H^+\to tb$~\cite{ATLAS:2016qiq} or $H^+\to \tau\nu$, which are suppressed or even zero in our scenario and hence not good signatures.

\subsection{Neutral Scalars}

The neutral scalars $H$ and $A$ have large couplings to $bs$, but also the far easier to detect muon coupling exists. For $A$, the branching ratio is however very small,
\begin{align}
\BR (A \to \mu^+ \mu^-) \simeq 2\times 10^{-4} c_{\beta\alpha}^4 \frac{1 - \BR_h^{bs}}{\BR_h^{bs}} \left(1 - \frac{m_H^2}{m_h^2}\right)^2 ,
\end{align}
especially in the region $m_H\sim m_h$ where the $h\to bs$ branching ratio $\BR_h^{bs}$ is largest. As a result, the best search channel is typically $A\to bs$. The same is true for $H$ in the limit $c_{\beta\alpha} \ll 1$, although a sizeable $c_{\beta\alpha}$ can lead to a large $H\to b\bar{b}$. With essentially only a large $bs$ coupling, $A$ can be produced at the LHC via the strange-quark sea, e.g.~$s g \to b A$, followed by $A\to bs$ or $A\to \mu\mu$. Similar channels have been discussed in the past, see for example Refs.~\cite{Dawson:2004nv,Altmannshofer:2016zrn}. For $H$, the $c_{\beta\alpha}$-suppressed gluon or vector-boson-fusion channels become available too, allowing for a search analogous to $h\to bs$.

A particularly interesting, albeit also $c_{\beta\alpha}$-suppressed, decay channel for $H,A$ is $H,A\to \gamma\gamma$. Recent $\sqrt{s}=\unit[13]{TeV}$ CMS limits for this signature can be found in Ref.~\cite{CMS:2017yta}, which also shows a small ($2.9\sigma$ local, $1.5\sigma$ global) excess around $m \simeq \unit[95]{GeV}$. This would be an interesting value for $m_H$, as it can lead to $\BR(h\to bs)\sim 20\%$ (Fig.~\ref{fig:mHvsBRbs}). With the couplings at hand, the cross section $pp\to H\to \gamma\gamma$ is simply too small for realistic values of $c_{\beta\alpha}$. However, the discussion so far assumed that all other entries $\epsilon^f_{ij}$ except $\epsilon^d_{23,32}$ are zero. Introducing extra couplings, in particular $\epsilon^u_{33}$, enhances both the gluon-fusion $H,A$ production as well as the $H,A$ branching ratio into $\gamma\gamma$ since $H,A$ with a mass of \unit[95]{GeV} cannot decay into two top quarks. In order to keep $h\to\gamma\gamma$ close to SM value, one needs $c_{\beta\alpha}\ll 1$, which in turn gives $m_A\simeq m_H$ due to Eq.~\eqref{eq:mA_kills_Bmix}. Therefore, the CMS diphoton excess would have to be interpreted as two unresolved peaks from $gg\to A/H\to\gamma\gamma$. Since the total signal corresponds approximately to the expected signal strength of an SM-like Higgs boson~\cite{CMS:2017yta} each boson should reproduce approximately half of the expected SM signal. Nevertheless, if one aims at large rates of $h\to bs$, very large values of $\epsilon^u_{33}$ will be required to obtain the desired $\gamma\gamma$-signal. We will leave a detailed discussion of this for future work.

\section{Different choice of basis}
\label{sec:different_basis}

\begin{table}
\begin{tabular}{|c|rrrrrr|}
\hline
$\rm{Type}$& ${c_y^d}$&${c_y^u}$&${c_y^\ell }$&${c_{\tilde \varepsilon }^d}$&${c_{\tilde \varepsilon }^u}$&${c_{\tilde \varepsilon }^\ell }$\\
\hline
I&$ {\cot \left( \beta  \right)}$&${\cot \left( \beta  \right)}$&${\cot \left( \beta  \right)}$&${ - \sin \left( \beta  \right)}$&${ - \sin \left( \beta  \right)}$&${ - \sin \left( \beta  \right)}$\\
II&${ - \tan \left( \beta  \right)}$&${\cot \left( \beta  \right)}$&${ - \tan \left( \beta  \right)}$&${\cos \left( \beta  \right)}$&${ - \sin \left( \beta  \right)}$&${\cos \left( \beta  \right)}$\\
X& ${\cot \left( \beta  \right)}$&${\cot \left( \beta  \right)}$&${ - \tan \left( \beta  \right)}$&${ - \sin \left( \beta  \right)}$&${ - \sin \left( \beta  \right)}$&${\cos \left( \beta  \right)}$\\
Y&$ { - \tan \left( \beta  \right)}$&${\cot \left( \beta  \right)}$&${\cot \left( \beta  \right)}$&${\cos \left( \beta  \right)}$&${ - \sin \left( \beta  \right)}$&${ - \sin \left( \beta  \right)}$\\
\hline
\end{tabular}
\caption{Relation between the parameters $\epsilon_{ij}^{f}$ of the Higgs basis and the new free parameters $\tilde{\epsilon}^{f}_{ij}$:  $\epsilon_{ij}^{f}=c_{y}^{f}y^f_{i}\delta_{ij}+\tilde{\epsilon}^{f}_{ij}/c_{\tilde{\varepsilon}}^{f}$. Here, $\tilde{\epsilon}^{f}_{ij}$ breaks the $\mathbb{Z}_2$ symmetry of the four 2HDMs with natural flavour conservation and induces flavour changing neutral currents.
}\label{tablebasis}
\end{table}

So far, we worked in the Higgs basis in which only one Higgs doublet requires a vacuum expectation value. However, this basis is not motivated by a symmetry and allows for generic large and potentially dangerous flavour violation, while the type-I, II, X and Y models posses a $\mathbb{Z}_2$ symmetry ensuring natural flavour conservation (see Ref.~\cite{Branco:2011iw} for an overview). Therefore, let us consider these models but allow for a breaking of this $\mathbb{Z}_2$ symmetry such that flavour changing Higgs couplings are possible. In Tab.~\ref{tablebasis} we give the relation between the couplings $\epsilon_{ij}^{f}$ defined in the Higgs basis and the quantities $\tilde{\epsilon}^{f}_{ij}$ which break the $\mathbb{Z}_2$ symmetry of the four 2HDMs with natural flavour conservation. Our new free parameters which induce flavour-changing neutral Higgs couplings are now $\tilde{\epsilon}^{f}_{ij}$ instead of ${\epsilon}^{f}_{ij}$. $\tan (\beta)$ corresponds as always to the ratio of the two vacuum expectation values.

First of all, we can rule out the type-II as well as the type-Y model since they lead to large effects in $b\to s\gamma$ and direct LHC searches, leading to stringent lower bounds on the masses of the additional scalars.

\subsection{Type-I Model}
Concerning $B_s$--$\overline{B}_s$ mixing the analysis remains unchanged compared to the one in the Higgs basis. For $B_s\to\mu^+\mu^-$ we can set $\tilde{\varepsilon}_{22}^{\ell}=0$, the condition to cancel $C_{S}-C^{\prime}_{S}$ reads
\begin{align}
m_{H}^{2}=\tan(\alpha)t_{\beta\alpha}m_{h}^2\,,
\end{align}
and $\tilde{\varepsilon}^d_{23,32}$ have to be chosen as
\begin{align}
\tilde{\varepsilon}_{23}^{d*}+\tilde{\varepsilon}_{32}^{d}=-\frac{4G_{F}^{2}m_{W}^{2} v C_{A}^\text{SM}}{\pi^{2}}\frac{m_b+m_s}{M_{B_{s}}^{2}}\frac{\tan(\beta)\sin(\alpha)}{c_{\beta\alpha}}m_{h}^{2}\,,
\end{align}
such that after destructive interference the SM result is recovered. Another possibility to avoid effects in $B_{s}\to\mu^{+}\mu^{-}$ is to choose $\tilde{\varepsilon}_{22}^{\ell}$ such that $C_{S}-C^{\prime}_{S}=0$, i.e.
\begin{align}
\tilde{\varepsilon}_{22}^{\ell}=-\frac{m_{\mu}}{v}\frac{\sin({\alpha})s_{\beta\alpha}m_{h}^{2}-\cos(\alpha)c_{\beta\alpha}m_{H}^{2}}{s_{\beta\alpha}^2 m_{h}^{2}+c_{\beta\alpha}^{2}m_{H}^{2}} \,.
\end{align}
This leads to the additional condition
\begin{align}
\tilde{\varepsilon}_{23}^{d*}+\tilde{\varepsilon}_{32}^{d}=\frac{4G_{F}^{2}m_{W}^{2} v C_{A}^\text{SM}}{\pi^{2}}\frac{m_b+m_s}{M_{B_{s}}^{2}} \frac{\sin(\beta)}{c_{\beta\alpha}s_{\beta\alpha}}\frac{m_{h}^{2}m_{H}^{2}}{m_{h}^{2}-m_{H}^{2}}\,.
\end{align}
Again, just like in the analysis in the Higgs basis, the effect in $B_s\to\mu^+\mu^-$ can be avoided if the neutral CP even scalars are degenerate in mass. Also concerning the anomalous magnetic moment of the muon one cannot expect a sizable effect due to the lack of any enhancement of the Higgs couplings to fermions. 

\subsection{Type-X Model}
Again, concerning $B_s$--$\overline{B}_s$ mixing the analysis remains unchanged compared to the one in the Higgs basis. For $B_s\to\mu^+\mu^-$ we can set $\tilde{\varepsilon}^{\ell}_{22}=0$, which requires
\begin{align}
m_{H}^{2}=-\cot(\alpha)t_{\beta\alpha}m_{h}^{2}
\end{align}
to get $C_{S}-C^{\prime}_{S}=0$. In addition,
\begin{align}
\tilde{\varepsilon}_{23}^{d*}+\tilde{\varepsilon}_{32}^{d}=\frac{4G_{F}^{2}m_{W}^{2} v C_{A}^\text{SM}}{\pi^{2}}\frac{m_b+m_s}{M_{B_{s}}^{2}}\frac{\cos(\alpha)}{c_{\beta\alpha}}m_{h}^{2}\,,
\end{align}
is needed. In the case of $\tilde{\varepsilon}^{\ell}_{22}\neq0$, the cancellation conditions read
\begin{align}
\tilde{\varepsilon}_{22}^{\ell}=\frac{m_{\mu}}{v}\frac{\cos(\alpha)s_{\beta\alpha}m_{h}^{2}+\sin(\alpha)c_{\beta\alpha}m_{H}^{2}}{s_{\beta\alpha}^2 m_{h}^{2}+c_{\beta\alpha}^2 m_{H}^{2}}
\end{align}
and
\begin{align}
\tilde{\varepsilon}_{23}^{d*}+\tilde{\varepsilon}_{32}^{d}=\frac{4G_{F}^{2}m_{W}^{2} v C_{A}^\text{SM}}{\pi^{2}}\frac{m_b+m_s}{M_{B_{s}}^{2}} \frac{\sin(\beta)}{c_{\beta\alpha}s_{\beta\alpha}}\frac{m_{h}^{2}m_{H}^{2}}{m_{H}^{2}-m_{h}^{2}}\,.
\end{align}
Here, in principle large effects in the anomalous magnetic moment of the muon are possible if $\tan(\beta)$ is large. However, in this case $B_s\to\mu^+\mu^-$ enforces $m_h\simeq m_H$ which leads simultaneously to a cancellation in the Barr--Zee contributions rendering the effect small again.

We find the following relations between the type I and the type X 2HDM in the case of $\tilde{\varepsilon}_{22}^{\ell}=0$
	\begin{align}
	\frac{m_{H}^{2}\big|_{\rm{Type}-I}}{m_{H}^{2}\big|_{\rm{Type}-X}}&=-\tan^{2}(\alpha) \,,\\
	\frac{\tilde{\varepsilon}_{23}^{d*}+\tilde{\varepsilon}_{32}^{d}\big|_{\rm{Type}-I}}{\tilde{\varepsilon}_{23}^{d*}+\tilde{\varepsilon}_{32}^{d}\big|_{\rm{Type}-X}}&=-\tan(\beta)\tan(\alpha)\,,
	\end{align}
	while in the case $\tilde{\varepsilon}^{\ell}_{22}\neq 0$ we obtain
	\begin{align}
	\frac{\tilde{\varepsilon}^{\ell}_{22}\big|_{\rm{Type}-I}}{\tilde{\varepsilon}^{\ell}_{22}\big|_{\rm{Type}-X}}&=\frac{m_{H}^{2}-\tan(\alpha)t_{\beta\alpha}m_{h}^{2}}{t_{\beta\alpha}m_{h}^{2}+\tan(\alpha)m_{H}^{2}} \,,\\
	\frac{\tilde{\varepsilon}_{23}^{d*}+\tilde{\varepsilon}_{32}^{d}\big|_{\rm{Type}-I}}{\tilde{\varepsilon}_{23}^{d*}+\tilde{\varepsilon}_{32}^{d}\big|_{\rm{Type}-X}}&=1\,.
	\end{align}

\section{Conclusion and Outlook}
\label{sec:conclusion}

The discovery of the Higgs boson has opened up new channels to search for flavour-violating processes. A comparison of $h\to f_i f_j$ with low-energy flavour observables is inherently model-dependent and thus difficult to assess in an effective-field-theory framework. In this article we have shown explicitly how the $h\to bs$ branching ratio can be enhanced to nearly arbitrary levels in a generic 2HDM while keeping other processes such as $B_s\to\mu^{+}\mu^{-}$, $B\to X_s \gamma$ and $B_s$--$\overline{B}_s$ mixing essentially at their SM values. Of course, this requires some tuning in the mass spectrum (new neutral scalars with masses similar to the SM Higgs) and couplings of the new scalars, but illustrates the importance of flavour-changing Higgs decays as a complementary probe of new physics. We strongly encourage dedicated experimental searches for $bs$ resonances.

Other rare or forbidden Higgs decays~\cite{Harnik:2012pb} can be analysed in a similar way within the 2HDM with generic Yukawa couplings:
\begin{itemize}
	\item $h\to bd$: Here the analogy with $h\to sd$ is straightforward, i.e.~the same conditions for the cancellations in flavour observables are required. However, the parameters must be adjusted even more precisely such that large decay rates can be possible.
	\item $h\to ds, uc$: Here the experimental problem of tagging light flavour makes it very hard to distinguish such modes from $h\to qq$ or $h\to gg$. Anyway, $\epsilon^{q}_{12,21}$ is stringently constrained from Kaon or $D$--$\bar D$ mixing. This bound can be avoided in the same way as the $B_s$--$\bar B_s$ mixing bound studied here. However, an even more precise cancellation would be required and bounds from $D\to\mu\nu$ and $K\to\mu\nu$ become relevant.
	\item $h\to \tau\mu$: Thanks to the former CMS excess in $h\to\tau\mu$~\cite{Khachatryan:2015kon}, many analyses already exist for this channel, showing that sizable rates are in fact possible, not only in the SM effective field theory with dimension-six operators but also in UV complete models (see for example Refs.~\cite{Campos:2014zaa,Heeck:2014qea,Sierra:2014nqa,Crivellin:2015mga,Dorsner:2015mja,Altmannshofer:2015esa,Altmannshofer:2016oaq,Herrero-Garcia:2016uab}). 
	\item $h\to \tau e$: Here the situation is very much like in the case of $h\to \tau\mu$ since the experimental bounds from $\tau\to e\gamma$ and $\tau\to e\mu\mu$ are comparable to the corresponding $\tau\to \mu$ processes.
	\item $h\to e\mu$: Obtaining large rates for $h\to\mu e$ is very difficult, not only because of the stringent bounds from $\mu\to e\gamma$ but also because of $\mu\to e$ conversion, where in a 2HDM~\cite{Crivellin:2014cta} an accurate cancellation among all the couplings to quarks would be required. 
\end{itemize} 
For a recent discussion of flavour violation involving the top quark see Ref.~\cite{Gori:2017tvg}.
Finally, $H,A\to\gamma\gamma$ in our model is particularly interesting in light of the CMS excess at \unit[95]{GeV}. By adjusting $\epsilon^u_{33}$ one can account for the measured signal since it only affects the effective coupling to gluons and photons but does not open up other decay channels.

\vspace{2ex}
\section*{Acknowledgements}
We thank Michael Spira for bringing the CMS $\gamma\gamma$ excess at \unit[95]{GeV} to our attention. J.H.~is a postdoctoral researcher of the F.R.S.-FNRS. The work of A.C.~and D.M.~is supported by an Ambizione Grant of the Swiss National Science Foundation (PZ00P2\_154834).

\appendix

\section{\texorpdfstring{Formulae for $b\to s\gamma$}{Formulae for b to s gamma}}
\label{sec:b_to_s_gamma_formulae}

Using the effective Hamiltonian
\begin{align}
H_\text{eff}^{b \to s\gamma } =  - \frac{{4{G_F}}}{{\sqrt 2 }}{V_{tb}}V_{ts}^ \star \sum\limits_i {{C_i}} {\mkern 1mu} {O_i} + \hc
\end{align}
with
\begin{align}
{O_7} &= \frac{e}{{16{\pi ^2}}}{m_b}\bar s{\sigma ^{\mu \nu }}{P_R}b \, {F_{\mu \nu }}\,,\\
{O_8} &= \frac{{{g_s}}}{{16{\pi ^2}}}{m_b}\bar s{\sigma ^{\mu \nu }}{T^a}{P_R}b \, G_{\mu \nu }^a\,,
\end{align}
where $F$ ($G$) is the electromagnetic (gluon) field strength tensor, we get the following expressions for the Wilson coefficients
\begin{widetext}
	\begin{align}
	C_{7,8}^{{H^+ }} &=\frac{1}{{4m_{{H^+ }}^2}}\frac{{\sqrt 2 }}{{4{G_F}{\lambda _t}}}\sum\limits_{j = 1}^3\left( {  \frac{{{m_{{u_j}}}}}{{{m_b}}}\Gamma _{L,j2}^{q{H^+ }*}{\mkern 1mu} \Gamma _{R,j3}^{q{H^+ }}f_{7,8}^{}\left( {{y_j}} \right) +   \Gamma _{L,j2}^{q{H^+ }*}{\mkern 1mu} \Gamma _{L,j3}^{q{H^+ }}g_{7,8}^{}\left( {{y_j}} \right)} \right){\mkern 1mu} ,\\
	C_7^{H_k^0} &=\frac{1}{{36m_{H_k^0}^2}}\frac{{\sqrt 2 }}{{4{G_F}{\lambda _t}}}\sum\limits_{j = 1}^3 \left({\Gamma _{R,2j}^{d{\kern 1pt} H_k^0}\Gamma _{R,3j}^{dH_k^0 \star } + \frac{{{m_s}}}{{{m_b}}}\Gamma _{R,j2}^{dH_k^0*}\Gamma _{R,j3}^{d{\kern 1pt} H_k^0} - \frac{{{m_{{d_j}}}}}{{{m_b}}}\Gamma _{R,2j}^{dH_k^0}\Gamma _{R,j3}^{d{\kern 1pt} H_k^0}\left( {9 + 6\log \left( {\frac{{m_{{d_j}}^2}}{{m_{H_k^0}^2}}} \right)} \right)} \right) ,\\
	C_8^{H_k^0} &=- 3 C_7^{H_k^0}\,,
	\end{align}
\end{widetext}

		with $y_{j}=m_{q_{j}}^2/m_{H^+}^2$ and the loop functions
		\begin{align}
		f_7^{}({y}) &={\mkern 1mu} {\frac{{ - 5y^2 + 8{y} - 3 + \left( {6{y} - 4} \right)\ln {y}}}{{{{3\left( {{y} - 1} \right)}^3}}}} \,,\nonumber\\
		f_8^{}({y})& = {\mkern 1mu} \frac{{ - y^2 + 4{y} - 3 - 2\ln {y}}}{{{{\left( {{y} - 1} \right)}^3}}} \,,\nonumber\\
		g_7^{}({y})&={\frac{{ - 8y^3 + 3y^2 + 12{y} - 7 + \left( {18y^2 - 12{y}} \right)\ln {y}}}{{{{18\left( {{y} - 1} \right)}^4}}}}  \,,\nonumber\\
		g_8^{}({y})&= {\mkern 1mu} {\frac{{ - y^3 + 6y^2 - 3{y} - 2 - 6{y}\ln {y}}}{{{{6\left( {{y} - 1} \right)}^4}}}}\,.
		\end{align}
		Here $\lambda_t=V_{tb}V_{ts}^ \star$ and we used the couplings $\Gamma$ defined as
		\begin{align}
		-\mathcal{L}_{Y}=&\sum_{f=u,d,\ell}\sum_{k}\bar{f}_{j}\left(\Gamma_{R,ij}^{H_{k}^{0}*}P_{L}+\Gamma_{R,ji}^{H_{k}^{0}}P_{R}\right)f_{i} H_{k}^{0}\nonumber\\
		&+\sqrt{2}\left[\bar{u}_{j}\left(\Gamma_{L,ji}^{qH^{+}}P_{L}+\Gamma_{R,ji}^{qH^{+}}P_{R}\right)d_{i}H^{+}+\hc\right]\nonumber\\
		&+\sqrt{2}\left[\bar{\nu}_{j}\Gamma_{R,ji}^{\ell H^{+}}P_{R}\ell_{i}H^{+}+\hc\right] ,
		\label{eq:lagrangian_gamma}
		\end{align}
with $H^0_{1,2,3}= h,H,A$.
In order to compare with the couplings given in Eq.~\eqref{eq:yukawas}, see Tab.~\ref{Tab:Gamma_couplings}.

\begin{table}[b]
	\begin{tabular}{|c|cc|}
		\hline
		&$\Gamma^{H}_{L}$&$\Gamma^{H}_{R}$\\
		\hline
		$H_{0}$&$c_{\beta\alpha}y^{f}-s_{\beta\alpha}\varepsilon^{f\dagger}$&$c_{\beta\alpha}y^{f}-s_{\beta\alpha}\varepsilon^{f}$\\
		$h_{0}$&$s_{\beta\alpha}y^{f}+c_{\beta\alpha}\varepsilon^{f\dagger}$&$s_{\beta\alpha}y^{f}+c_{\beta\alpha}\varepsilon^{f}$\\
		$A_{0}$&$-i\eta_{f}\varepsilon^{f\dagger}$&$i\eta_{f}\varepsilon^{f}$\\
		$H^{+}$&$-\varepsilon^{u\dagger}V$&$V\varepsilon^{d}$\\
		$H^{+}$&$0$&$\varepsilon^{\ell}$\\
		\hline
	\end{tabular}
	\caption{The couplings of the interaction Lagrangian in Eq.~\eqref{eq:lagrangian_gamma} in terms of $y^{f}$ and $\varepsilon^{f}$ in Eq.~\eqref{eq:yukawas}.}
	\label{Tab:Gamma_couplings}
\end{table}

\bibliographystyle{utcaps_mod}
\bibliography{BIB}

\end{document}